# A High Position Resolution X-ray Detector: an "Edge on" Illuminated Capillary Plate Combined with a Gas Amplification Structure

C. Iacobaeus[1], T. Francke[2], B. Lund-Jensen[3], J. Ostling[4], P. Pavlopoulos[5,6], V. Peskov[5], F. Tokanai[7]

[1]Karolinska Institute, Stockholm, Sweden
[2]XCounter AB, Danderyd, Sweden
[3]Royal Institute of Technology, Stockholm, Sweden
[4]Stockholm University, Stockholm, Sweden
[5]Leonard de Vinci University, Paris, France
[6]CERN, Geneva, Switzerland
[7]Yamagata University, Yamagata, Japan

*Abstract*

We have developed and successfully tested a prototype of a new type of high position resolution hybrid X-ray detector. It contains a thin wall lead glass capillary plate converter of X-rays combined with a microgap parallel-plate avalanche chamber filled with gas at 1 atm.

The operation of these converters was studied in a wide range of X-ray energies (from 6 to 60 keV) at incident angles varying from 0-90º. The detection efficiency, depending on the geometry, photon's energy, incident angle and the mode of operation, was between 5-30% in a single step mode and up to 50% in a multi-layered combination. Depending on the capillary's geometry, the position resolution achieved was between 50-250 μm in digital form and was practically independent of the photon's energy or gas mixture.

The usual lead glass capillary plates operated without noticeable charging up effects at counting rates of 50 Hz/mm$^2$, and hydrogen treated capillaries up to 10$^5$ Hz/mm$^2$.

The developed detector may open new possibilities for medical imaging, for example in mammography, portal imaging, radiography (including security devices), crystallography and many other applications.

# I. INTRODUCTION

Gaseous detectors always attracted users by their simplicity, high avalanche gains and position resolutions. However, the stopping power of gases is low and this restricts the application of the gaseous detectors. On the other hand, solid–state detectors which have high stopping powers, are very expensive and cannot provide high avalanche gains.

The ideal approach would be of course to develop a detector which combines the advantages of high stopping power and a high position resolution typical for solid-state detectors with high avalanche multiplications offered by gaseous detectors.

In the past we have demonstrated that CsI coated capillary plates (CPs) can be used as converters of soft X-rays combined with gas multiplication structures [1]. This hybrid detector offered an excellent position resolution of 30-50 μm; however, the efficiency for the hard X-rays was around 1% only [2].

In this work we tried another approach: using "edge-on" illuminated, thin-wall CPs as X-ray converters. As will be shown below, this allowed us to achieve both high position resolutions and high efficiencies even for hard X-rays.

This paper describes our first results obtained in this promising direction.

## II. STUDY OF THIN WALL CPs AS X-RAY CONVERTORS

### A. Experimental Set Up

Our experimental set up is shown schematically in Fig. 1. Essentially it contains a test chamber and an X-ray gun with a tungsten anode. In most measurements a filter and a collimator were placed in front of the chamber. The test chamber was installed on a special step motor controlled table which allows for 3D alignment to take place with up to a few μm of accuracy. The table also allowed chamber rotation for up to 90°. Behind the test chamber a PM with a NaI scintillator was installed. When the chamber was removed from the table, this detector was used for the measurements of the X-ray intensity mainly in the range of photon energies of 10-60 keV. For low intensity beams we also used a CZT detector. The measurements with the CZT detector show that a heavily filtered X-ray beam had a sharp peak in the spectrum at a photon energy $E_{max}$ slightly below the voltage $V_g$ applied to the X-ray gun. Thus after heavy filtering one can obtain an almost monochromatic beam with characteristics energy $E_{max} \sim V_g$. For some control measurements radioactive X-ray sources were used: $^{55}$Fe, $^{109}$Cd and $^{241}$Am. Inside the test chamber various CPs and gas multiplication structures could be installed (see Fig. 2). The CPs tested were made of lead glass and had thicknesses of 0.8-1 mm; the holes had diameters of 12, 30 and 100 μm and the wall thickness was of 2.5, 6 and ~20 μm respectively. Two to three cm above the CP a mesh was installed. If necessary, the gap between the mesh and the CP could be used as a drift region for primary electrons created there by X-rays. In some measurements a readout plate was also placed 0.4 mm below the CP. It was a ceramic plate with Cr strips of a 50 μm pitch. A voltage of 1-2 kV could be applied between the CP and the readout plate. This made it possible for avalanche multiplication to be achieved in this region. Depending on the measurements one of the two readout plates was used. The first one had all of the strips interconnected together to a single charge sensitive amplifier or to a Kethley picoampermeter. This allowed one to use this plate either for counting rate measurements (counting rate $N_{CP}$ of avalanches produced in the amplification region) or for various measurements of the current produced by the X-rays. The second readout plate had 120 strips in the central region of the plate connected to an ASIC. This allowed for accurate position resolution measurements. In some measurements a cathode micromesh was installed between the anode of the CP and the readout plate supported by "fishing" lines 100 μm thick served as spacers [3]. In fact it was a combination of the CP with a home made MICROMEGAS.

B. *Measurements*

The CPs' efficiency η for photon energies of 30-60 keV was determined as

$$\eta = N_{CP}/N_D \qquad (1)$$

where $N_D$ is the counting rate from the PM or CZT for a heavily filtered X-ray beam. The cross checks of the efficiencies were also made with several X-ray radioactive sources: $^{55}$Fe, $^{109}$Cd and $^{241}$Am. Current measurements mentioned above were performed in order to independently verify the measurements made with the PM and the CZT detectors for the photos with energy of ≤30 keV. This method of the efficiency measurements is illustrated by Fig. 3. The upper curve represent the current (measured on the top electrode of the CP) vs. the voltage between the drift mesh and the CP's top. One can see that with the voltage the current first sharply increased and then reached the plateau (the saturated value of $I_s$). The lowest curve shows the current measured on the CP's bottom electrode vs. the voltage across the CP. In these measurements the drift electrode was grounded. The behavior of this current is different from the previous one: first it increased, then reaches the saturated value $I_{CP}$ (corresponding to full extraction of electrons from the capillary holes-see [1] for more details) and then increased again (due to the avalanche gain in the CP). These measurements allowed us to determine the ratio of $\xi = I_s/I_{CP}$. The efficiency was determined as

$$\eta = \{1/\xi(1-e^{-kx})\} \qquad (2)$$

where k is a linear absorption coefficient in the gas for the X-rays with the given energy $E_v$ and x - the thickness of the drift region. Such current measurements were done for each gas mixtures used: $Ar+10\%CH_4$, $Xe+10\%CH_4$, $Kr+20\%CO_2$ or $He+10\%CH_4$ at pressures of 1 atm. The characteristic energy $E_v$ was determined from the convolution of the measured X-ray spectrum with the absorption curve of the gas. Typically $E_v$ was only slightly below $V_g$.

C. *Summary of results*

The operation of the converters was investigated in a wide range of X-ray energies (from 6 to 60 KeV) and at incident angles (from 0 to 90º). Some results of efficiency measurements are presented in Table 1. As one can see from this data the highest sensitivity was achieved with thin-walled capillaries (holes' diameter of 12-30 μm) in an "edge on" illuminated mode. We also tested multilayer CP converters; in this case the efficiency could reach up to 50% (see [4] for details).
A position resolution was measured with a collimated X-ray beam of 30 μm in diameter. At a gas gain of 10 a position resolution of 150-250 μm in digital form was achieved (see [4]). At gains of more than 1 000 the position resolution improved to about 50 μm [4].

## III. AN X-RAY DETECTOR BASED ON "EDGE- ON" ILLUMINATED CPs

Based on the results described above we have built a simplified prototype of an X-ray scanner which is shown schematically in Fig. 4. A lead glass capillary plate 0.8 mm thick and with hole diameters of 12 μm was used in the "edge on" illuminated mode as an X-ray converter. The primary electrons extracted from capillary holes trigger avalanches in the gap between the CP bottom and the readout plate. The measurements were performed at gas gains of about $3·10^4$. As an example Fig. 5 shows an image of two wires placed 50 μm apart. One can see that at the position resolution of 50 μm was achieved in digital mode.

## III.IV. DISCUSSION AND CONCLUSIONS

The performed studies and obtained images demonstrate the potentials of the new detector. High efficiency ( up to 37%), high gas gains (> 1000) and high position resolutions (~50 μm) were achieved at the same time. Note that usual lead glass capillary plates can operate without noticeable charging up effects at counting rates of 50-100 Hz/mm$^2$ and hydrogen treated capillaries–up to $10^5$ Hz/mm$^2$ [5].

The developed detectors may open new possibilities for medical imaging, for example in mammography (scanning mode), radiography (including security devices) and crystallography. Thin wall multi-layered CPs irradiated perpendicular to their surfaces could also be used as X-ray converters for a portal imaging device [5].

## IV.V. REFERENCES:

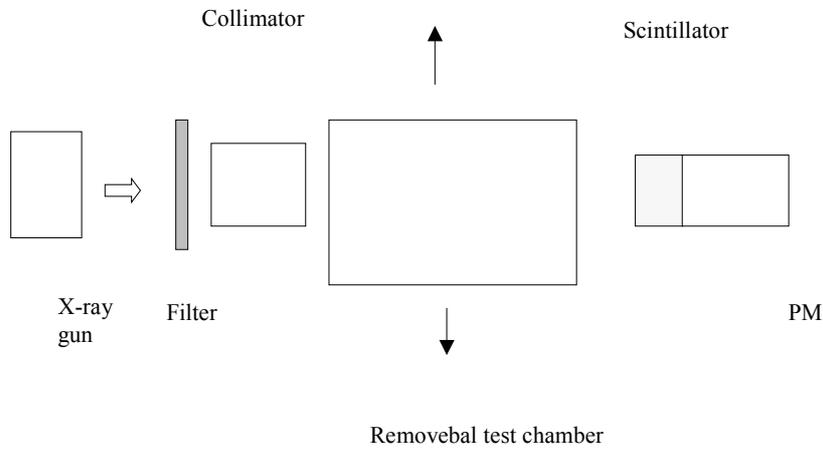

Fig. 1. A schematic drawing of the experimental set up for measurements of the position resolution and the efficiency of the hybrid gaseous detectors.

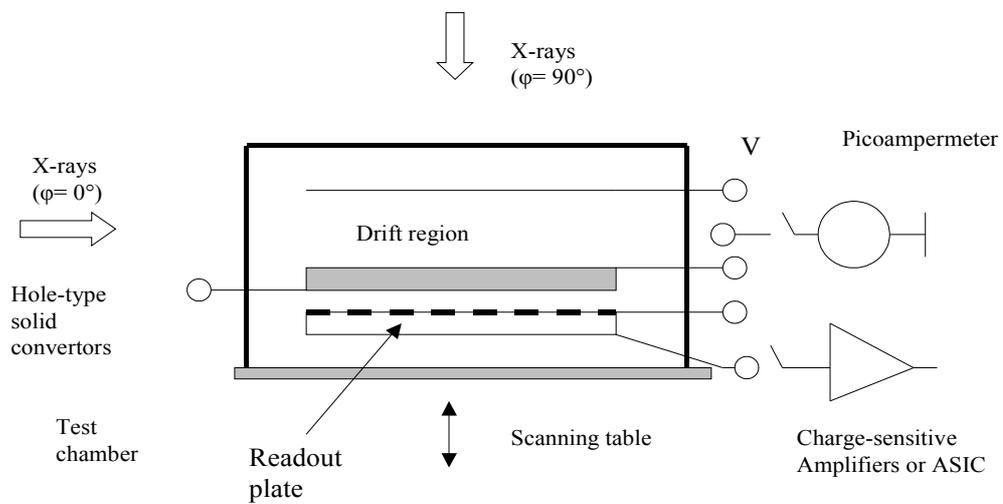

Fig. 2. The schematic drawing of the inner structure of the test chamber and the associated equipment.

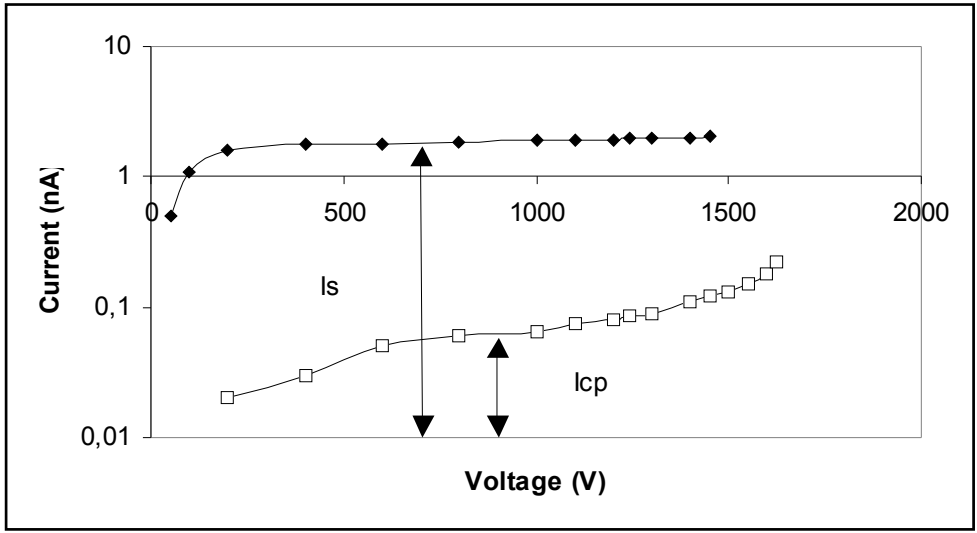

Fig. 3. Currents measured in the drift space (upper curve) due to the X-rays interaction with gas and from the CP (lower curve) due to the X-rays interactions with capillary walls. $V_g$=15 keV. Gas mixture Kr+20%$CO_2$.

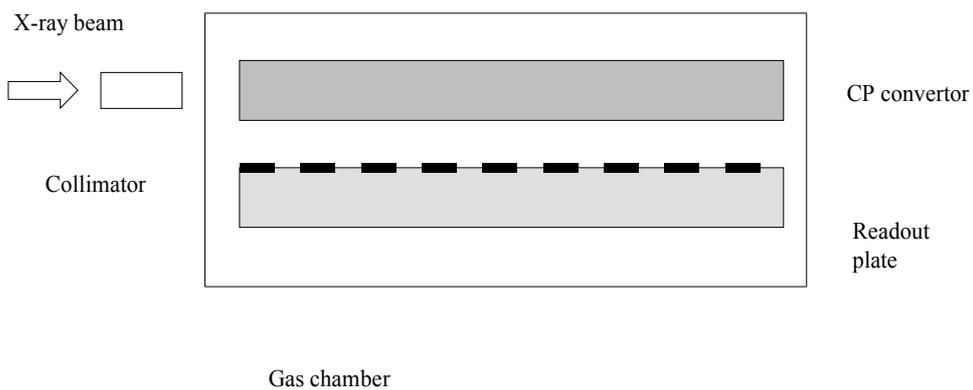

Fig. 4. A schematic drawing of the X-ray scanner prototype.

| V. Capillary type | VI. Efficiency η(%) for different angles φ (data in bracket-X-ray tube voltage in kV): | | | |
|---|---|---|---|---|
| Hole's diameter, μm: | φ=0° | φ=30° | φ=45° | φ=90° |
| 12 | η=2.1 (22)<br>η= 12.8(35)<br>η=37 (60) | η=9.3(30) | η=7.5 (30) | η=3.2(10)<br>η= 6.3 (30)<br>η=3.75 (60) |
| 30 | η=5.2 (35)<br>η=23 (60) | η=3.8 (35) | η=3.2 (35) | η=2 (10)<br>η=2.8 (35)<br>η~1(60) |
| 100 | η= 0.1 (10)<br>η= 0.8(30)<br>η= 6(60) | η=1.8 (10) | η=2.2(10) | η=2.6 (10)<br>η=1.5(30)<br>η=0.8(60) |

Table 1: Some results of the CPs efficiency measurements. Note that 22 and 60 keV photons were produced by [109]Cd and [241]Am sources. Photons of other energies were from the heavily filtered X-ray gun radiation.

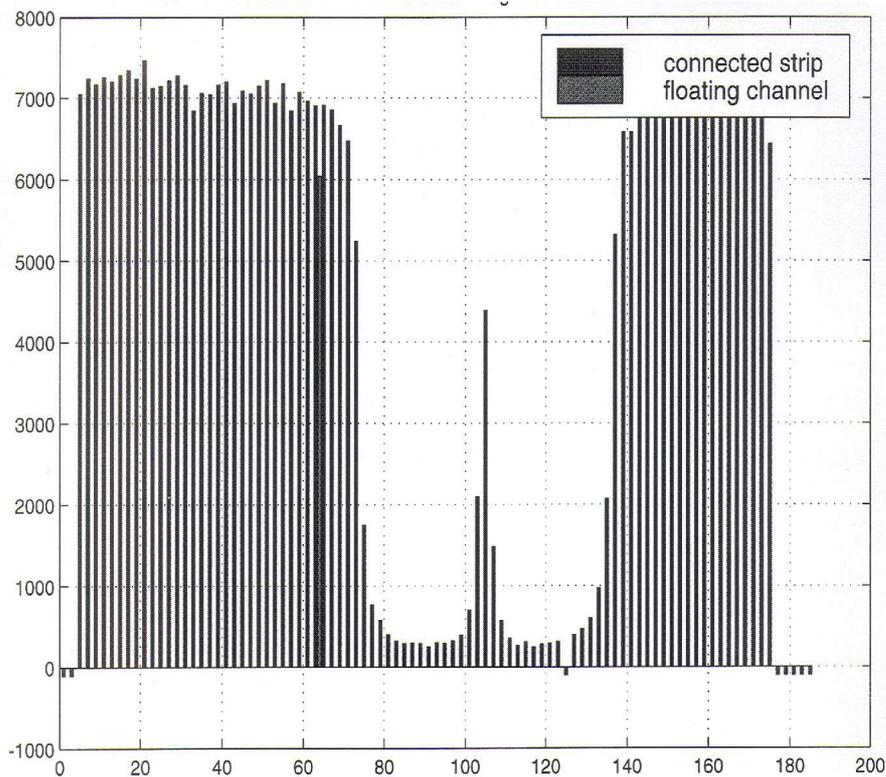

Fig.5 On line digital image of two wires 50 μm apart from each other. Vertical scale- number of counts, horizontal scale- channel number